\documentclass{osa-article}

\journal{osajournal}

\usepackage{xfrac}
\usepackage{graphicx}
\usepackage{dcolumn}
\usepackage{bm}
\usepackage{epstopdf}
\usepackage{verbatim}
\usepackage{soul}
\usepackage{ulem}
\usepackage{xcolor}

\DeclareMathOperator{\sinc}{sinc}
\newcommand{\kk}{\mathbf{k}}
\newcommand{\qq}{\mathbf{q}}
\newcommand{\ww}{\omega}

\begin{document}

\title{Spatial and spectral characterization of photon pairs at telecommunication-wavelength from type-0 spontaneous parametric down-conversion}

\author{Evelyn A. Ortega,\authormark{1,2*} Jorge Fuenzalida, \authormark{1,2,3} Mirela Selimovic,\authormark{1,2$\dagger$} Krishna Dovzhik,\authormark{1,2} Lukas Achatz,\authormark{1,2} Sören Wengerowsky,\authormark{1,2,4} Rodrigo F. Shiozaki,\authormark{5} Sebastian Philipp Neumann,\authormark{1,2} Martin Bohmann,\authormark{1} and Rupert Ursin\authormark{1,2}}

\address{\authormark{1}Institute for Quantum Optics and Quantum Information - IQOQI Vienna, Austrian Academy of Sciences, Boltzmanngasse 3, 1090 Vienna, Austria.\\
\authormark{2}Vienna Center for Quantum Science and Technology (VCQ), Vienna, Austria.\\
\authormark{3}Current address: Fraunhofer Institute for Applied Optics and Precision Engineering IOF, Albert-Einstein-Str. 7, 07745 Jena, Germany.\\
\authormark{4}Current address: ICFO-Institut de Ciencies Fotoniques, The Barcelona  Institute of Science and Technology, 08860 Castelldefels (Barcelona), Spain.\\
\authormark{5}Departamento de F\'isica, Universidade Federal de S\~{a}o Carlos, Rodovia Washington Lu\'is, km 235—SP-310, 13565-905 S\~{a}o Carlos, SP,Brazil.}

\email{\authormark{*}evelynacunaortega@gmail.com}
\email{\authormark{$\dagger$}mirela.selimovic@univie.ac.at}

\begin{abstract}
The thorough characterization of entangled-photon sources is vital for their optimal use in quantum communication.
However, this task is not trivial at telecommunication wavelengths.
While cameras and spectrometers are well developed for visible and near-infrared spectra, that does not apply in the mid-infrared range.
Here we present a spatial and spectral characterization of photon pairs emitted in a type-0 phase-matched spontaneous parametric down-converted source.
We experimentally show how these photon properties are modified by the crystal temperature. 
This parameter allows easy modification of photon-pair properties to fit novel multiplexing schemes based on only one entanglement photon source.
Our results pave the way for the optimal design and use of spatial and spectral properties of quantum-correlated photon pairs at telecommunication wavelengths.
\end{abstract}

\date{\today}

\section{Introduction}

Nonclassical light is at the heart of a large number of applications in  quantum communication \cite{PanQKD2020}, quantum imaging \cite{gilaberte2019perspectives, fuenzalida2022res}, and quantum computation \cite{zhong2020quantum}.
For generating nonclassical light, the most well-studied and widely used process is spontaneous parametric down-conversion (SPDC) \cite{klyshko1969scattering,burnham1970observation}.
The SPDC photons can feature quantum correlations in a variety of degrees of freedom such as polarization \cite{sansa2022visible}, time \cite{Franson89,carvacho2015postselection}, space \cite{WALBORN201087,OAM2001}, or frequency \cite{frec2009}.
In particular, high-dimensional entangled systems have been promising in multiplexing schemes \cite{Chen_2009, Pseiner_2021, Puttnam:21} and high-dimensional encoding \cite{erhard2020advances,Achatz_2022}.

Down-converted photons are spatially and spectrally correlated due to the momentum and energy conservation during the SPDC process.
These correlations depend on the pump beam's properties and the phase-matching condition associated with the nonlinear crystal properties \cite{Anwar2021}.
In periodically-poled crystals, the photon-pair emission characteristics can be tuned by varying the crystal temperature \cite{Fejer92}.
In particular, it can be used to change the photon pair's spectral and spatial properties, i.e., from collinear to non-collinear propagation as well as from degenerate to non-degenerate cases.
The crystal temperature is an easily accessible parameter in entangled photon-pair source setups since it is adjusted by an electronic controller.
Therefore, manipulating and optimizing spatial and spectral correlations by tuning the crystal temperature could be the way to tailor the SPDC photons for desirable applications without changing optical components in the source configuration.

Spatial correlations have been thoroughly investigated for type-I and type-II phase-matched SPDC, including their dependence on the pump waist \cite{PhysRevA.83.033837, bennink11}, crystal length \cite{Ram_rez_Alarc_n_2013}, and phase-matching conditions \cite{PhysRevA.80.022307, laweberly2004}.
Type-0 phase-matched SPDC sources have an exceptionally high brightness, and a broad spectrum \cite{Steinlechner:14} which can be exploited to generate photon pairs in the telecommunication band for distribution through optical fibers \cite{ghatak2008}.
These features allow for a wide range of integration into existing communication infrastructure \cite{Wengerowsky6684, neumann2022quapital} and the implementation of multi-user networks \cite{Joshi2020}.
However, for photon pairs at telecommunication wavelength generated in type-0 phase-matched SPDC, spatial correlations characterization based on the crystal temperature has not been implemented yet.
In this spectral region, there are technical difficulties in collecting and analyzing the spatial and spectral distributions due to the lack of compact and efficient single-photon cameras and spectrometers.

In this paper, we study the crystal-temperature dependency of the transverse momentum and spectrum of photon pairs generated by type-0 phase-matched SPDC around $1550\,$nm.
We implement a versatile experimental setup where the phase-matching is changed through temperature control in a magnesium-oxide-doped periodically poled lithium niobate (MgO:ppLN) crystal.
This analysis includes the collinear and non-collinear emission scenarios.
We evaluate momentum correlations of the photon-pair emission through a stepwise scanning at the far-field plane.
Additionally, we quantify their spatial entanglement using the so-called Schmidt number.
Finally, we measure the down-converted photons' spectral characteristics by a tunable wavelength filter at different crystal temperatures.
Our results provide a comprehensive and precise characterization of spatial and spectral properties of photons pairs, which can help develop novel quantum communication schemes based on the high-dimensional entangled systems.

The paper is structured as follows:
in Sec. \ref{Sec:theory}, we recall the theory of type-0 phase-matched SPDC process, placing special emphasis on their temperature dependence.
In Sec. \ref{Sec:Exp_setup}, our experimental setups are explained, including the photon-pair source as well as the measurement of spatial and spectral correlations.
Our results are presented and discussed in Sec. \ref{Sec:Results}.
We summarize and conclude our findings in Sec. \ref{Sec:conclusion}.

\section{Theory of type-0 phase matching SPDC}
\label{Sec:theory}

We briefly recall the theoretical description of type-0 SPDC sources and compute the characteristics of photon pairs generated in a MgO:ppLN crystal.
In SPDC, a pump photon ($p$) of angular frequency $\ww_p$ propagating through a nonlinear material is annihilated, creating two correlated photons labeled signal ($s$) and idler ($i$) of angular frequencies $\ww_s$ and $\ww_i$, respectively.
Let us assume that the pump photon propagates along the $z$-axis of a periodically-poled crystal with a poling period of length $\Lambda$.
The SPDC process must satisfy the conservation of energy and momentum which results in $\omega_p=\omega_s+\omega_i$, and the quasi phase-matching (QPM) condition $\bigtriangleup\kk=\kk_{p}-\kk_{s}-\kk_{i}-\kk_{m}$ \cite{Fejer92}.
Where $\kk_p$, $\kk_s$, and $\kk_i$ are the wave vectors of the respective pump, signal and idler photons.
And $\kk_{m}$ is the grating wave vector defined by $|\kk_{m}|=2\pi/\Lambda$.
Thus the bi-photon mode function in the momentum representation is given by \cite{WALBORN201087}
\begin{equation}
\label{eq:amplitudSPDC}
\Phi (\qq_s,\qq_i,\omega_{s},\omega_{i})\propto E_{p}\left(\qq_s+\qq_i\right) \sinc\left(\frac{\mathrm L}{4\kk_p}\left\vert \qq_s-\qq_i\right\vert ^{2}+\varphi\left(\mathrm T,\lambda\right)\right).
\end{equation}
Here, $\mathrm L$ is the crystal length, and $\qq_s$ ($\qq_i$) is the transverse wave vector of the signal (idler) photon, i.e., $\qq_j=(q_{x_j},q_{y_j})$.
The angular spectrum of the pump beam $E_{p}\left(\qq_s+\qq_i\right)$ typically is represented as a Gaussian function.
The $\sinc$ function, called the phase-matching function, plays a crucial role in the spectral and spatial properties of the photon pairs.
The collinear phase mismatch parameter $\varphi\left(\mathrm T,\lambda\right)$ relates the possible wavelength combinations of the three photons ($\lambda_p,\lambda_i,\lambda_s$) and the crystal temperature $\mathrm T$ through the refractive indices defined by the Sellmeier equation associated with the crystal \cite{Gayer2008}.

The momentum conservation in the process guarantees that signal and idler photons are emitted with opposite transverse wave vectors, which means that $\qq_s=-\qq_i$.
Hence, in the far-field plane of the crystal, we observe a photon pair distribution at opposite transverse positions, i.e., they are anti-correlated.
Due to the rotational symmetry in the emission, it is sufficient only to consider one transverse coordinate to calculate the emission angle relative to the principal axis of the crystal. 
Thus, for the $x$-axis
\begin{equation}
\label{eq:angle}
\mathbf\theta_{j}=\arctan{\frac{\left\vert \mathbf{q}_{x_{j}}\right\vert}{\left\vert\mathbf{k}_{z_j}\right\vert}},
\end{equation}
with $\vert\mathbf{k}_{z_j}\vert$ being the longitudinal wave-vector component.
Where the collinear emission corresponds to $\theta_{j}\approx0$, i.e., signal and idler photons are emitted in the same direction as the pump photon.
\begin{figure}[h!]
\centering
\includegraphics[width=0.9\textwidth]{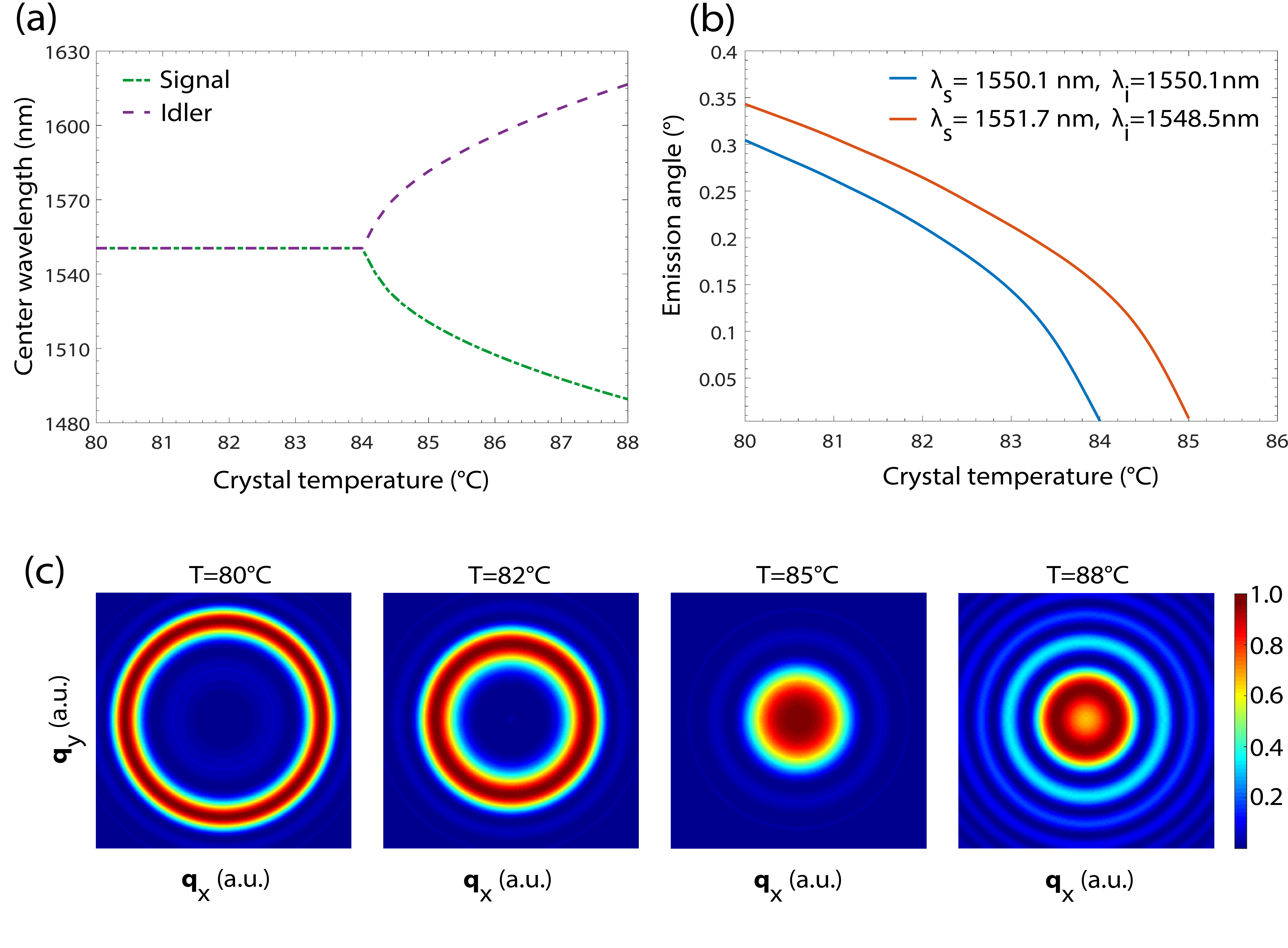}
\caption{(a) Numerically calculated center wavelength for photons emitted from MgO:ppLN crystal as a function of its temperature.
(b) Calculated emission angle of the photon pairs as a function of the crystal temperature. 
For illustration, we selected the degenerate case (blue) and a non-degenerate one (red).
(c) Representation of spatial intensity distribution of the photons at the far-field plane with $\lambda_{i}=1551.72$~nm and $\lambda_{s}=1548.52$~nm at four crystal temperatures: $80$, $82$, $85$ and $88^\circ$C.
}
\label{fig:theory}
\end{figure}

To compare the theoretical predictions with our experimental results, we consider the following input parameters: a pump wavelength of $\lambda_p$ $\sim 775$~nm, crystal parameters $\mathrm L=40$~mm, and $\Lambda=19.2$~$\mu$m.
We compute the spectral properties by the square modulus of the phase-matching function and the emission angle by Eq.~(\ref{eq:angle}), displayed in Fig.~\ref{fig:theory}(a) and \ref{fig:theory}(b), respectively.
Figure~\ref{fig:theory}(a), for crystal temperatures between $80^\circ$C to $84^\circ$C shows spectral degeneracy, i.e., $\lambda_{i}=\lambda_{s}$.
For higher temperatures up to $88^\circ$C photons pairs show non-degenerate wavelengths with tunability over $1490-1617$~nm.
Another spectral characteristic that has shown temperature dependence is the spectral bandwidth \cite{Steinlechner:14}.
The full width at half maximum (FWHM) bandwidth is determined by crystal length and the photon pairs' group velocity mismatch, resulting in a broad spectrum close to the degenerate case.
Here the FWHM is about $60$~nm and decreases in non-degenerate wavelength until $\sim12$~nm at $\mathrm T=90^\circ$C for signal and idler separately.
In Fig.~\ref{fig:theory}(b), we observe the relation between emission angle and wavelength of photon pairs depending on the crystal temperature. 
Different photon-pair wavelengths selection affects the emission angle at the same temperature.
The blue line represents the degenerate case with $\lambda_{s}=\lambda_{i}= 1550.1$~nm, showing collinearity at $\mathrm T \approx 84^\circ$C.
The red line represent a non-degenerate case with $\lambda_{i}=1551.72$~nm, $\lambda_{s}=1548.52$~nm, which results in a collinear cone at $\mathrm T \approx 85^\circ$C.
In Fig.~\ref{fig:theory}(c), we can see the QPM transition of signal and idler photons from a non-collinear to a collinear propagation from Eq.~(\ref{eq:amplitudSPDC}).

\section{Experimental implementation}
\label{Sec:Exp_setup}

\subsection{Photon-pair source}

The experimental setup of the type-0 phase-matched SPDC source is shown in Fig.~\ref{fig:setup}.
A $775$~nm continuous-wave (cw) pump laser is coupled into a single-mode ﬁber (SMF) and sent through a polarizing beam splitter (PBS) and a half-wave plate (HWP) to adjust the pump beam's polarization for which the type-0 phase-matching was designed.
The collimated pump beam is focused with a plano-convex lens $\text{L}_0$ with focal length $f_0=250$~mm at the center of a $40$~mm long MgO:ppLN crystal with a poling period of $\Lambda=19.2$~$\mu$m. 
The resulting pump waist in the center of the crystal was $\sim 80$~$\mu$m.
The nonlinear crystal was mounted in an oven allowing a homogeneous temperature distribution inside of the oven where the crystal temperature was set with a precision of $\pm 0.1^\circ$C.
The emitted SPDC photons were separated from the pump beam with a dichroic mirror (DM) and spectrally cleaned with a long-pass filter (LPF).
The lens $\text{L}_1$ is located at a focal distance $f_1=200\,$mm from the center of the crystal, performing a Fourier transform of its transverse plane at the focal distance $f_1$.

\subsection{Spatial-correlation measurement}

\begin{figure}[h!]
\centering
\includegraphics[width=0.9\textwidth]{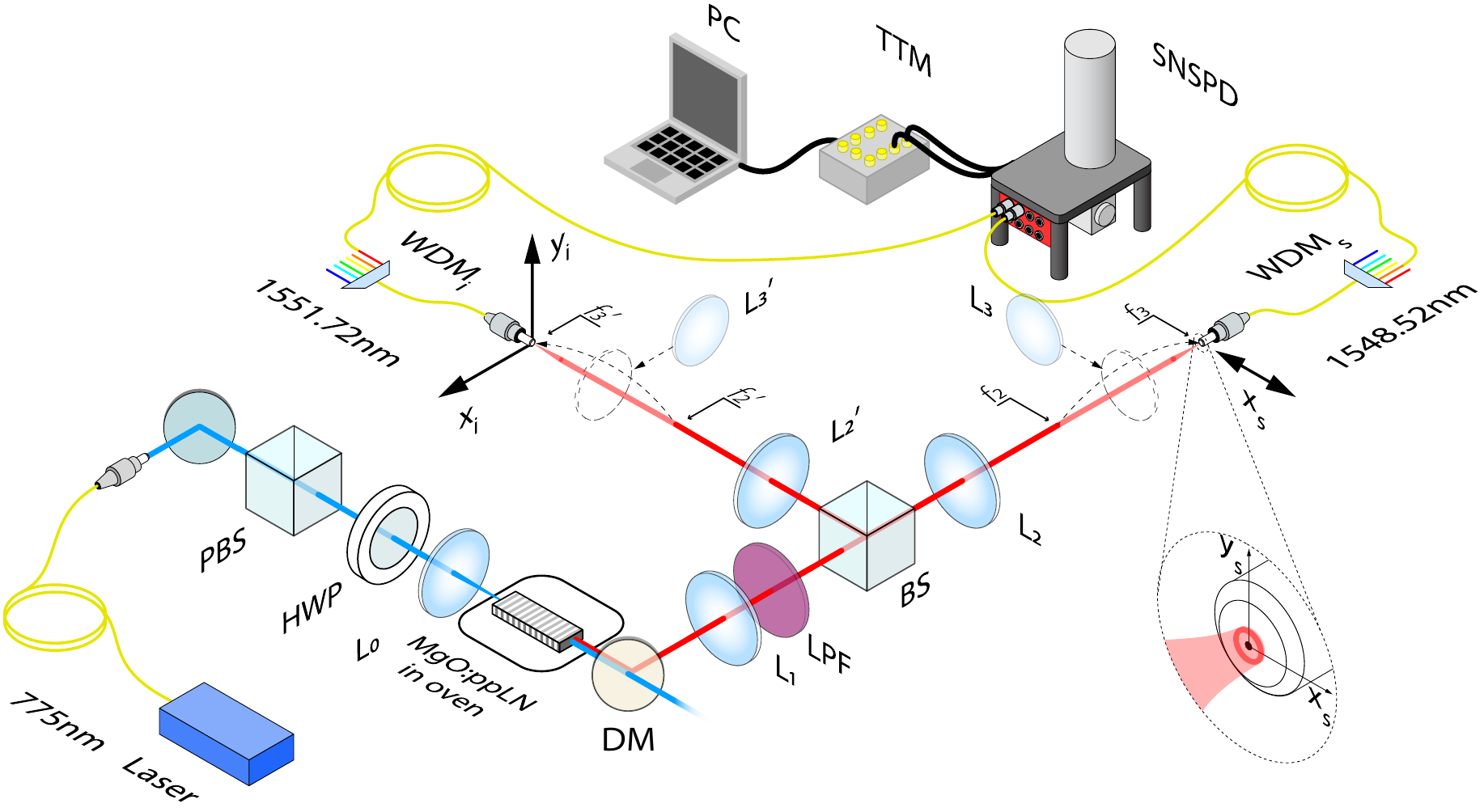}
\caption{Experimental setup of the type-0 phase-matched SPDC source and experimental spatial-correlation measurement.
A 40 mm-long MgO:ppLN crystal was mounted in an oven heated to a specific temperature.
It was pumped with a $775$~nm cw laser (blue line) to emit SPDC photon pairs at telecommunication wavelength (red lines).
The lens $\text{L}_1$ is located at focal distance $f_1=200\,$mm from the center of the crystal.
Two optical configurations are used to implement either near- or far-field measurements.
For the near-field measurement, $\mathrm{L}_2$ ($\text{L}_2'$) was located at a distance $f_1+f_2$ ($f_1+f_2'$) from L1 and the fiber translation stages were placed at the focal distance $f_2=f_2'=150$~mm from $\mathrm{L}_2$ ($\text{L}_2'$).
For the far-field measurement, $\mathrm{L}_3$ ($\text{L}_3'$) was mounted at the focal distance $f_2+f_3$ ($f_2'+f_3'$) from $\mathrm{L}_2$ ($\text{L}_2'$) and the fiber translation stage was moved back at the focal distance $f_3=f_3'=4.5$~mm from $\mathrm{L}_3$ ($\text{L}_3'$).
In both scenarios, the photons were detected using superconducting nanowire single-photon detectors (SNSPD).
}
\label{fig:setup}
\end{figure}

To analyze the spatial correlations of the photon pairs emitted by our source and its dependence on the crystal temperature, we implemented scanning measurements in the near-field (NF) and far-field (FF) of the signal and idler mode (see Fig.~\ref{fig:setup}).
At the output of the source, the signal and idler photons were split probabilistically on a 50:50 beam-splitter (BS).
For recording the far-field plane, we implemented an imaging system $2f_{2}/2f_{3}$ ($2f_2'/2f_3'$) with the lenses $\text{L}_2$ and $\text{L}_3$ ($\text{L}_2'$ and $\text{L}_3'$) with focal lengths of $f_2=f_2'=150$~mm and $f_3=f_3'=4.5$~mm.
This configuration results in a magnification factor of $\text{M}_{\text{FF}}=0.03$.
For recording the near-field plane, the crystal's center is imaged with a $2f_1/2f_2$ ($2f_1/2f_2'$) system using the lenses $\text{L}_1$ and $\text{L}_2$ ($\text{L}_2'$) in each arm.
This configuration results in a magnification factor of $\text{M}_{\text{NF}}=0.75$.

In the near- and far-field configurations, the photon pairs were collected in their respective detection planes by two telecommunication wavelength SMF mounted on translation stages.
Due to the symmetry in the SPDC photons emission, we performed for the idler a plane scan $(x_i,y_i)$ and for the signal just a linear scan along the $x$-axis $(x_i,y_i=0)$.
The scan was performed over $100\,\mu$m in steps of $10\,\mu$m, which were chosen following the mode field diameter of the SMF $\sim10.4~\mu$m.
For each crystal temperature, we recorded $21^3=9261$ data points, corresponding to the three-dimensional measurement settings $(x_i, y_i; x_s)$.
The wavelength of the signal and idler were filtered by appropriate wavelength-division-multiplexing (WDM) channels with a spectral FWHM of $0.6$~nm.
We chose the wavelength channels centered at $\lambda_{i}=1551.72\,$nm and $\lambda_{s}=1548.52\,$nm for a detailed analysis as they are close to the central wavelength.
The output ports of the WDMs were connected to superconducting nanowire single-photon detectors (SNSPD) with $\sim80\%$ efficiency and dark-count rates of $\sim10^2$ Hz.
Detection events were digitized and time-stamped with the aid of a time-tagging module (TTM).
For each crystal temperature, we collected at the position of the fibers $1\,$s the single and coincidence counts between signal and idler.
The coincidence counts were identified when both detectors registered photons within a $200\,$ps time window.

\subsection{Spectral-correlation measurement}

The second parameter in consideration is the spectral distribution of the photon pairs as a function of the crystal temperature.
The spectral characterization of the entangled photon-pair source was achieved via spectral separation of the SPDC photons by setting up a tunable wavelength filter utilizing a reflective diffraction grating.

\begin{figure}[h!]
\centering
\includegraphics[width=0.9\textwidth]{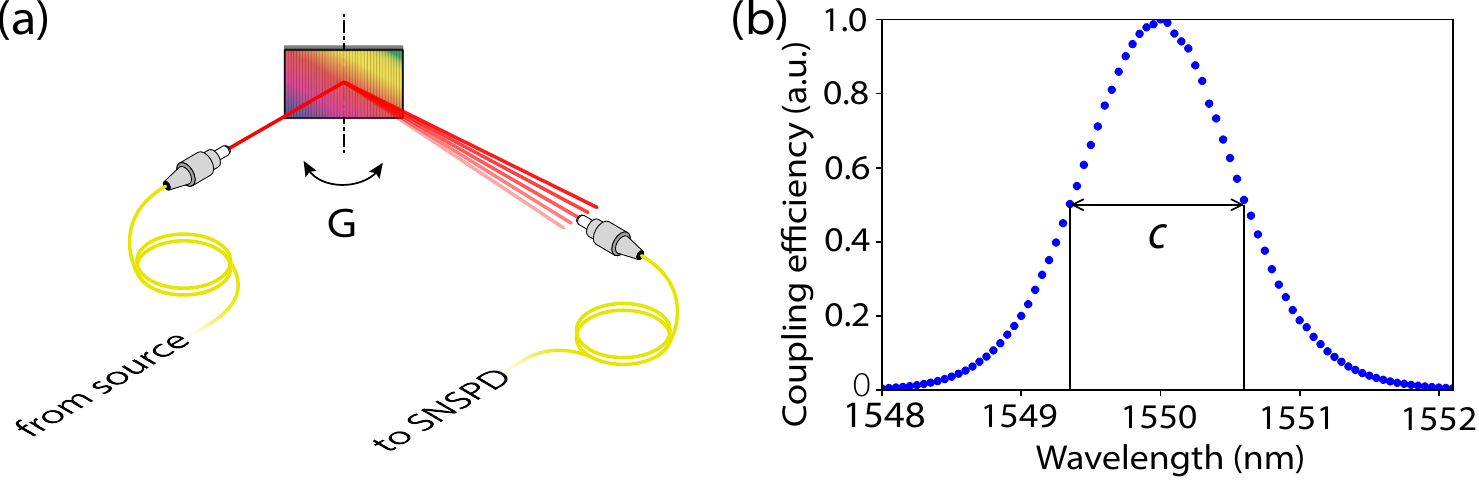}
\caption{(a) Experimental setup of the single-mode optical spectrometer optimized for telecommunication wavelengths.
The signal of the entangled photon-pair source is coupled out from the optical fiber and led onto a blazed grating G.
Subsequently, the beam is partially coupled into another single-mode fiber that is connected to a SNSPD.
(b) The experimental curve for the coupling of a wavelength tunable laser at fixed grating position is shown.
The full width at half maximum of the Gaussian-shaped curve was determined to be $\text{C} = 1.25\pm0.05$~nm and can be interpreted as the smallest perceptible difference in the wavelength measurable with our filter setup.
}
\label{fig:setupspectrum}
\end{figure}

In this case, after lens $\mathrm{L}_1$ and LPF (see Fig. \ref{fig:setup}), both signal and idler photons were coupled into a SMF and directed to the input of the tunable wavelength filter setup (see Fig.~\ref{fig:setupspectrum}(a)).
After the SMF, the SPDC photons propagated towards a dispersive grating in free space. 
By reflection on the periodic grating, the SPDC photons were diffracted and fanned out, resulting in a spatially continuous wide signal. 
An SMF at the output acted as a wavelength filter of the fanned-out broad signal. 
The grating was positioned onto a rotation stage and by rotation the wavelength section that is led to the detector is altered.

The angular dispersion of the grating is $1.46$~nm/mrad which resulted in a wavelength section width of $1.25$~nm that is coupled into the output SMF of the tunable wavelength filter setup.
It defines the width of the wavelength bins that are detected for each rotation step.
The width was determined by connecting and varying the wavelength of a tunable laser with fixed grating position resulting in Fig.~\ref{fig:setupspectrum}(b).
FWHM of the Gaussian-shaped curve is $\text{C}= 1.25$~nm, which can be interpreted as the resolution of our spectrum analyzer.

The output SMF of the tunable wavelength filter was connected to an SNSPD channel.
By step-wise rotation, a full spectrum scan of the SPDC photons was recorded.
The measured property is the count rate of single photons per wavelength bin emerging from the SPDC process.
Each rotation step was collected and registered with the TTM over $10$ s.

\section{Results}
\label{Sec:Results}

\subsection{Spatial correlations of type-0 phase-matched SPDC}

\subsubsection{Scan results}
\label{sec:scan}

We recorded the transverse momentum of the signal and idler photons by scanning its far-field planes for crystal temperatures ranging from $80^{\circ}$C to $90^{\circ}$C in steps of $0.5^{\circ}$C.
To obtain the single count rate for every $(x_i, y_i)$ position, we average the recorded counts of the idler mode over all positions of the $(x_s, y_s=0)$ motor stage, which corresponds to an integration time of $21$~s at each point.
Furthermore, coincidence count rates were obtained for linear scans, i.e., $(x_i, y_i=0)$ and $(x_s, y_s=0)$.

\begin{figure}[h!]
\centering
\includegraphics[width=0.85\textwidth]{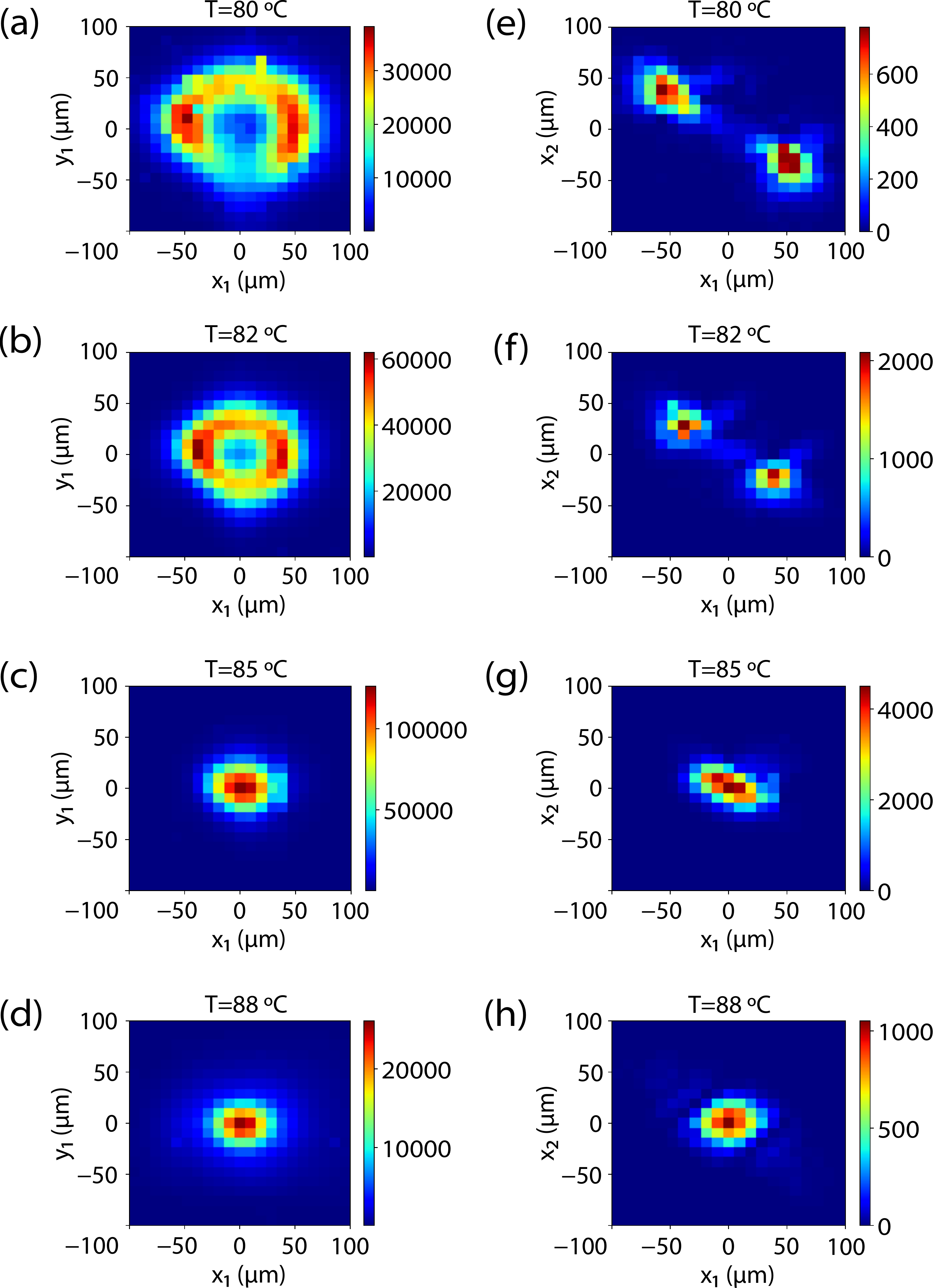}
\caption{Measurement of the photon spatial distribution in the far-field for the crystal temperatures $80$, $82$, $85$, and $88^{\circ}$C from top to bottom.
(a-d) Average single counts per second for every $(x_i, y_i)$.
(e-f) Coincident counts per second for every $(x_i, x_s)$, when fiber positions $(y_i, y_s)$ were at $0~\upmu$m.}
\label{fig:rings}
\end{figure}

Figure~\ref{fig:rings} shows the significant experimental results of the transverse plane scan in far-field over all crystal temperature.
Figures~\ref{fig:rings}(a-d) show the single count rates for crystal temperatures $80$, $82$, $85$ and $88^{\circ}$C, respectively.
Figures~\ref{fig:rings}(e-h) depict the coincidence count rates at the same temperatures.
It should be noted that the single count rates correspond to the far-field intensity profile of the photons. 
In the same plane, the coincidence count rates evidence the anti-correlations in transverse momenta. 
All experimental data are normalized to their maximum value since the fiber coupling and SPDC efficiency change for each crystal temperature.

These experimental results demonstrate that the variation in the crystal temperature modifies the spatial distributions of the single and coincidence count rates in far-field.
Based on the crystal temperature dependence of the phase-matching function, the collinear case was found at $\mathrm{T}=85^{\circ}$C in Fig.~\ref{fig:rings}(c), where the photon pairs are distributed in a small central spot with the highest coincidence count rate in Fig.~\ref{fig:rings}(g).
Decreasing the temperature below the collinear case, i.e., $\mathrm{T}<85^{\circ}$C, yields to the SPDC ring-opening in Fig.~\ref{fig:rings}(a) and \ref{fig:rings}(b).
The radius of the SPDC ring increases while the radial thickness of the ring decreases accordingly.
In this case, the coincidence count rates are obtained by linear scans so that they are registered along the $x$-axis in opposite emission directions, as are shown in the non-collinear cases in Fig.~\ref{fig:rings}(e) and 4(f).
Alignment imperfections cause asymmetries in non-collinear cases when the SPDC ring expands.
Figure~\ref{fig:rings}(h) shows that for $\mathrm{T}>85^{\circ}$C, the efficiency of the nonlinear process decreases drastically.
This is due to the SPDC photon intensity distribution exhibiting as multiple secondary rings in Fig.~\ref{fig:rings}(d), not just in the central spot as in the collinear case.
The experimental results in Fig.~\ref{fig:rings} are qualitatively in good agreement with theoretical predictions depicted in Fig.~\ref{fig:theory}(c) at the same crystal temperatures.

\subsubsection{Schmidt-number estimation}
\label{sec:schmidt}

To quantify the spatial correlations of photon pairs as a function of the crystal temperature, we use the Schmidt number \cite{PhysRevLett.92.127903}.
We determine the Schmidt number of the entangled photon pairs by comparing the near- and far-field intensity distributions as introduced in \cite{PhysRevA.80.022307}\footnote{This method allows verifying entanglement by measuring one subsystem with the assumption that the bipartite state to be pure. The more entangled the bipartite state, the more incoherent are its one photon states. By measuring the overall degree of coherence of one subsystem one can then make assumptions about the overall quantum state.}.
This approach draws from classical coherence theory to estimate the Schmidt number $\mathrm K$ through
 \begin{equation}
 \mathrm K \approx \frac{1}{(2\pi)^2} \frac{\left[\int \text{I}_{\text{NF}}(\textbf{r})d\textbf{r}\right]^{2}}{\int \text{I}_{\text{NF}}^{2}(\textbf{r})d\textbf{r}} \times \frac{\left[\int \text{I}_{\text{FF}}({\boldsymbol{\textbf{q}}})d{\boldsymbol{\textbf{q}}}\right]^{2}}{\int \text{I}_{\text{FF}}^{2}({\boldsymbol{\textbf{q}}})d{\boldsymbol{\textbf{q}}}},
\label{eq:schmidt}
\end{equation}
where  $\text{I}_{\text{NF}}(\textbf{r})$ is the intensity in the near-field and $\text{I}_{\text{FF}}(\textbf{q})$ is the intensity in the far-field \cite{Achatz_2022, vanExter:09}.

\begin{figure}[h!]
\centering
\includegraphics[width=0.85\textwidth]{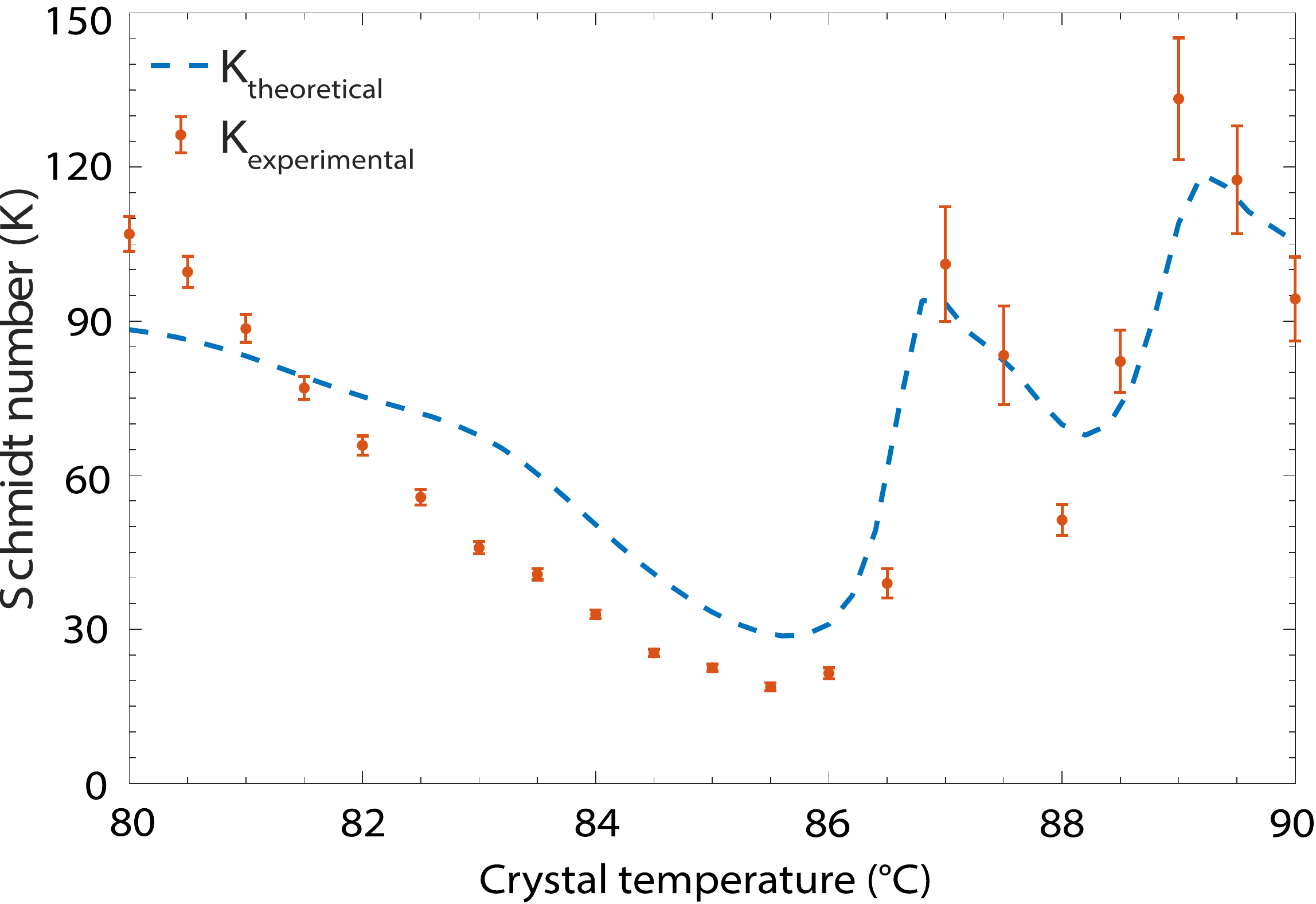}
\caption{Schmidt number while varying the crystal temperature for photon pairs generated with our type-0 phase-matched SPDC source.
The experimental (orange dots) and theoretical (dashed blue line) Schmidt number was obtained from the far-field and near-field intensities.
The theoretical prediction was obtained by considering the SPDC photon state given by Eq.~(\ref{eq:amplitudSPDC}) and its Fourier transform.
}
\label{fig:schmidt}
\end{figure}

The Schmidt numbers obtained from our experimental data for each crystal temperature are presented in Fig~\ref{fig:schmidt}.
Orange dots correspond to the experimental data, and error bars result from the Poissonian photon counting statistics and Gaussian error propagation.
We calculated K using Eq.~\eqref{eq:schmidt} with the theoretical description for the momentum representation given by Eq.~(\ref{eq:amplitudSPDC}) and the position representation is obtained through the Fourier transform of Eq.~(\ref{eq:amplitudSPDC}).
Our experimental results are in good agreement with theory prediction considering the non-collinear and collinear propagation by varying the crystal temperature.
Figure~\ref{fig:schmidt} shows that starting from $\mathrm{T}=80.0^{\circ}$C, the value of $\mathrm K$ decreases as the temperature increases until reaching its minimum value $\mathrm K\approx 18$ at $\mathrm{T}=85.5^{\circ}$C.
This point is close to the collinear case and marks an inflection point. 
If the temperature rises above $\mathrm{T}>85.5^{\circ}$C, $\mathrm K$ starts to increase irregularly with oscillations.
The multiple SPDC rings that can be recognized in its spatial distribution increase the spatial modes of the Schmidt decomposition (c.f. Eq.~(\ref{eq:amplitudSPDC})).
Although $\mathrm K$ is higher as the crystal temperature increases, the emission intensity drops rapidly.
This is in contrast to $\mathrm{T}<85.5^{\circ}$C, where the SPDC photons are emitted in only one ring that is opened up while the radial thickness of the ring decreases.
This corroborates that the amount of spatial entanglement could be increased by changing the crystal temperature.

\subsection{Temperature-dependent spectrum in type-0 phase-matched SPDC}

\begin{figure}[h!]
\centering
\includegraphics[width=0.9\textwidth]{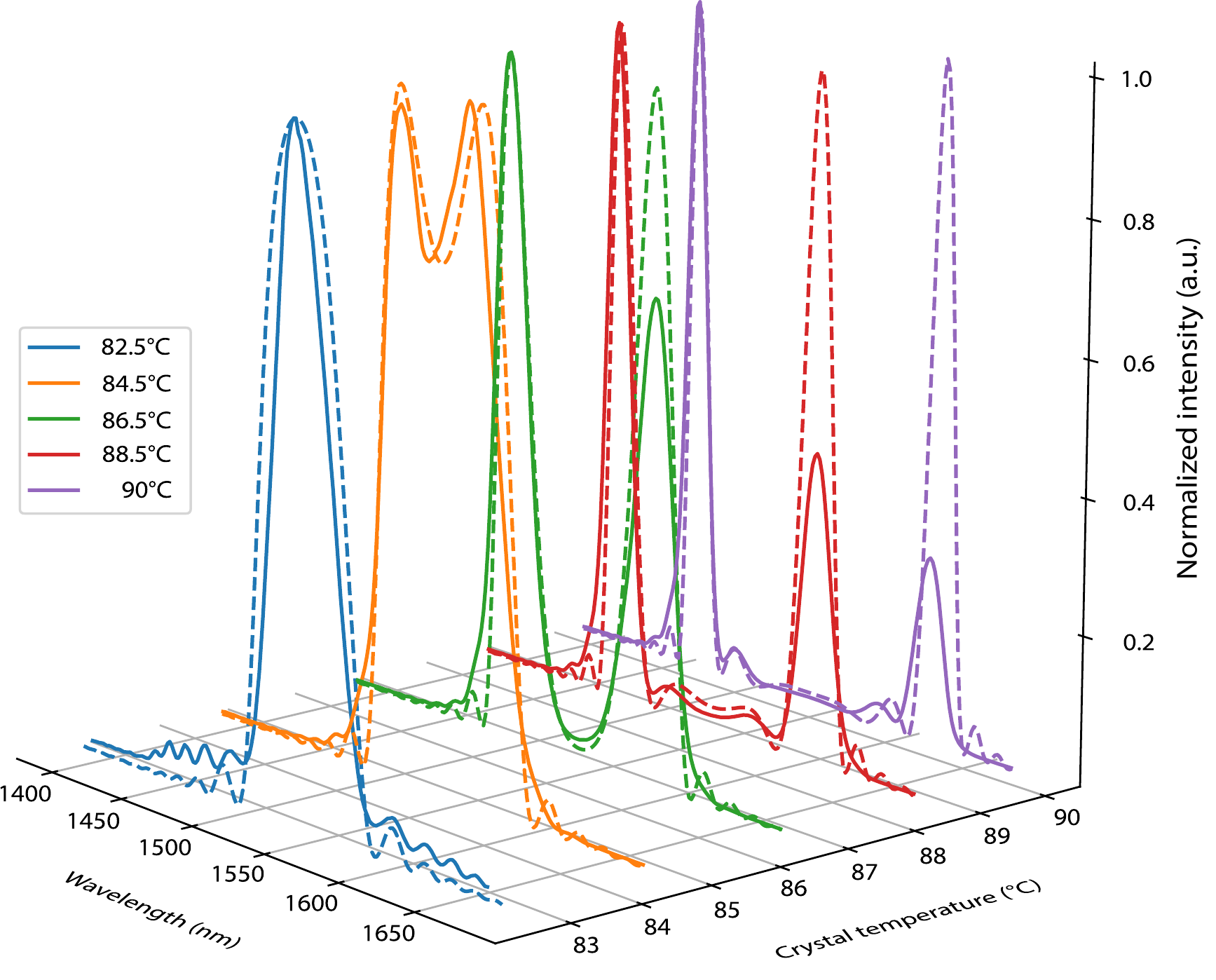}
\caption{Experimentally (solid lines) and theoretically calculated (dash lines) normalized emission spectra as a function of the wavelength of the type-0 entangled photon-pair source for different crystal temperatures.
The data was collected with the tunable wavelength filter setup.}
\label{fig:spectrum}
\end{figure}

We measured the spectrum of the entangled photon-pair source for different crystal temperatures in a range from $82.5^\circ$C to $90^\circ$C.
Figure~\ref{fig:spectrum} shows the theoretical and experimental normalized SPDC emission spectra at different crystal temperatures.
Solid lines represent the measured spectra of the photon pairs with crystal temperatures: $82.5$, $84.5$, $86.5$, $88.5$ and $90^\circ$C.
The dashed lines are the theoretically predicted curves for the temperatures of the Sellmeier equation for MgO:ppLN crystals \cite{Gayer2008}.
It can be observed how the SPDC spectra change from degenerate to non-degenerate wavelengths while increasing crystal temperature.
Also, we can distinguish the large bandwidth characteristic for type-0 phase-matched SPDC around the degeneracy temperature $\sim 84^\circ$C.
It is evident from Fig.~\ref{fig:spectrum} that the source produces degenerate photons at a spectral FWHM bandwidth of $\sim 40$ nm.
The bandwidths of the emission spectra decrease for higher temperatures.

Although the experimental spectra are in good agreement with the theoretically estimated spectra, for non-degenerate cases, we observed some asymmetry in the recorded intensities.
A plausible explanation for the observed asymmetry lies in the fact that photons with lower frequencies diverge more than its twin photons, resulting in a lower coupling efficiency into the single-mode mode fiber and, thus, lower count rates.
Another relevant factor that could contribute to the asymmetry is the wavelength dependence of the detector efficiency since the detector has an efficiency peak at $1550$~nm.
For higher wavelengths the efficiency decreases resulting into a lower detector efficiency.
The fact that the asymmetry in the measured intensities increase with increasing degeneracy is consistent with both explanations.

\section{Conclusion}
\label{Sec:conclusion}

We have investigated the spatial and spectral properties of photon pairs at telecommunication wavelength generated in a type-0 phase-matched SPDC source by varying its crystal temperature.
The spatial correlations were experimentally measured by a scanning approach, while spectral properties were measured with a single-mode optical spectrometer optimized for telecommunication wavelengths.
We further report a good agreement between experimental results and theoretical predictions.
As expected, different crystal temperatures allow significant changes between non-collinear to collinear emissions and degenerate to non-degenerate wavelength.
The crystal temperature is an accessible parameter in entangled photon-pair sources, and we have proven that it is an excellent tool to tailor them to different configurations.

Our results have practical importance in developing entanglement distribution schemes based on spatial and spectral correlations.
This is particularly interesting in photon pairs at telecommunication wavelengths that can be incorporated into deployed telecommunication infrastructures and integrated with different optical fibers.
Recent implementations of multicore fiber space-division \cite{MCF2022}, and entanglement distribution multiplexing \cite{Joshi2020, Pseiner_2021}, enhance the quantum channel performance in quantum communication protocols.
Combined with our results, these implementations enable novel communication schemes with multiple users selecting configurations according to the crystal temperature.
Our characterization provides a pathway for entanglement spatial and spectral distribution in quantum networks and brings more flexibility to entangled photon pair sources.

\section*{Funding}
This work was supported by the European Union’s Horizon 2020 programme grant agreement No.857156 (OpenQKD) and the Austrian Academy of Sciences.
\section*{Acknowledgments}
E.A.O. and J.F. acknowledge ANID for the financial support (Becas de doctorado en el extranjero “Becas Chile”/2016 – No. 72170402 and 2015 – No. 72160487).
L.A. acknowledges financial support from the EU project OpenQKD (Grant agreement ID: 85715).

\section*{Disclosures}
The authors declare no conflicts of interest.

\bibliography{Biblio}

\end{document}